\documentclass[31pt, oneside]{article}   	
\usepackage{geometry}                		
\usepackage{graphicx}				
\usepackage{amsmath}
\newtheorem{definition}{Definition}[section]

\usepackage{amssymb}
\usepackage{indentfirst}
\usepackage{authblk}

\title{Calculating worst-case Value-at-Risk prediction using empirical data under model uncertainty}
\author{Wentao Hu \thanks{Institute for Financial Studies and School of Mathematics, Shandong University, Jinan 250100, China}}
\date{July 31, 2019}

\begin{document}
\maketitle

\begin{abstract}
Quantification of risky positions under model uncertainty is of crucial importance from both viewpoints of external regulation and internal management. The concept of model uncertainty, sometimes also referred to as model ambiguity. Although we know the the family of models, we cannot precisely decide which one to use. Given the set $\mathcal{P}$, the value of the risk measure $\rho$ varies in a range over the set of all possible models. The largest value in such a range is referred to as a worst-case value, and the corresponding model is called a worst scenario. Value-at-Risk (VaR) has become a very popular risk-measurement tool since it was first proposed. Naturally, WVaR(worst-case Value-at-Risk) attracts the attention of many researchers. Although many literatures investigated WVaR, the implications for empirical data analysis remain rare. In this paper, we proposed a special model uncertainty market model to simply the $\mathcal{P}$ to a set contain finite number of probability distributions. The model has the structure of the two-layer mixed distribution model. We used change point detection method to divide the returns series and then used EM algorithm to estimate the parameters. Finally, we calculated VaR, WVaR(worst-case Value-at-Risk) and BVaR(best-case Value-at-Risk) for four financial markets and then analyzed their different performance.\par
Keywords: Value-at-Risk; worst-case; model uncertainty; empirical data; mixed distribution model
\end{abstract}

\section{Introduction}
	With the rapid development of the financial industry and the frequent emergence of financial crisis, risk management has become an important issue faced by company managers and government regulators.Value-at-Risk (VaR) has become a very popular risk-measurement tool since it was first proposed by J.P. Morgan. Let $X$ be the loss,
\begin{eqnarray}
VaR_{\alpha}(X) = inf\{x\in \mathbb{R}:P[X\leqslant x]> \alpha\} = F^{-1}_{X}(\alpha).
\end{eqnarray} 	
It has achieved the high status of being written into industry regulations. Simplicity is an important advantage. First of all, VaR is easy to understand. According to Duffie\cite{duffie1997}, VaR can be defined as For a given time horizon $T$ and a confidence level $\alpha$, VaR is the loss in market value that can only be exceeded with a probability of at most $1-\alpha$. That is ``We are $\alpha\%$ certain that we will not lose more than $V$ dollars in the next $N$ days.'' VaR is simply the $\alpha$ percentile of the loss distribution. Using a single number to describe complex financial risks can make the measurement of risk simple and intuitive. In addition, Kupiec\cite{Kupiec1995}, Christoffersen\cite{Christoffersen1998} and Hull\cite{Hull2012} indicated that VaR is easy to calculate and get back-test. However, Lo\cite{Lo1988} and  Leskow\cite{Leskow2001} claimed that VaR suffers from being unstable and difficult to work with numerically when losses are not ``normally" distributed--which in fact is often the case, because loss distributions tend to exhibit "fat-tails". Thus, Zangari\cite{Zangari1996} and Venkataraman\cite{Venkataraman1997} pointed out that VaR under normal assumption will lead to serious underestimation of the risk of extreme losses. \par
	Many researchers made efforts to solve this problem. Huschens\cite{Huschens1999}, Glasserman\cite{Glasserman2002} and Lin\cite{Lin2006} used multivariate t-distribution to amend the original assumption. Li\cite{Li1999}, Wilhelmsson\cite{Wilhelmsson2010}, Su\cite{Su2011} and Gabrielsen\cite{Gabrielsen2012} used statistics such as volatility, skewness and kurtosis to capture the extreme tail. Haas\cite{Haas2006} and Gebizlioglu\cite{Gebizlioglu2011} used Weibull distribution. Zangari\cite{Zangari1996}, Venkataraman\cite{Venkataraman1997} and Hull\cite{Hull1998} estimated VaR with Gaussian mixture. Zhang\cite{Zhang2005} uses an EM algorithm based on KS test to determine the number of component distributions. Billio\cite{Billio2000} and Kawata\cite{Kawata2007} used switching volatility model to describe the independence of data. For an up-to-date account of relative literature, some recent review papers Kuester\cite{Kuester2006}, Jorion\cite{Jorion2010}, Abad\cite{Abad2014} and Zhang\cite{Zhang2017} can be referred.\par 	
	However, quantification of risky positions held by a financial institution under model uncertainty is of crucial importance from both viewpoints of external regulation and internal management. Given some sources of uncertainty a Bayesian methodology, O'Hagan\cite{Hagan1994}, Bernardo\cite{Bernardo1994} and Alexander\cite{Alexander2012} assumed that VaR is described in terms of a set of unknown parameters. Bayesian estimates may be derived from posterior parameter densities and posterior model probabilities which are obtained from the prior densities via Bayes theorem, assuming that both the model and its parameters are uncertain. The problem has also been studied from a non-Bayesian point of view, such as Modarres\cite{Modarres2007}, Giorgi\cite{Giorgi2008} and Figlewski\cite{Figlewski2003}. As for VaR model, Alexander\cite{Alexander2012} shared ideas with the Bayesian approach. Jorion\cite{Jorion1996} and Talay\cite{Talay2002} investigated sampling error. 
	However, we remark that there is no consensus on the sources of model risk. The concept of model uncertainty, sometimes also referred to as model ambiguity. It was argued that data follow not a single distribution but rather a family of distributions. Although we know the the family of models, we cannot precisely decide which one to use.  Cont\cite{Cont2006} quantified the model risk of a complex product by the range of prices obtained under all possible valuation models. Many literatures studied the worst-case value of risk measures with given partial information. Very often, the problem of interest is of the following type: to find
\begin{eqnarray}
\text{sup}\ \rho(X),\ X\in \mathcal{P},
\end{eqnarray}
where $\rho$ is a risk measure, and the set $\mathcal{P}$ is a class of random variables with some given partial distributional information. Given the set $\mathcal{P}$, the value of the risk measure $\rho$ varies in a range over the set of all possible models. The largest value in such a range is referred to as a worst-case value, and the corresponding model is called a worst scenario. An early source is Royden\cite{Royden1953}. Kass\cite{Kaas1984}, Schepper\cite{Schepper2010}, Popescu\cite{Popescu2005}, He\cite{He2010} and many other literatures calculated the bound of $P(X\leqslant x)$ under partial information. Peng\cite{Peng2004,Peng2006,Peng2008,Peng2010} proposed sublinear expectation. The central concept of sublinear expectation theory is a family of distributions inherent in the data series. Artzner\cite{Artzner1999} proposed the concept of coherent risk measures, which can be viewed as a special instance of sublinear expectation. Although many literatures investigated worst-case value of risk measures, the implications for real data analysis remain rare. Peng\cite{Peng2018} proposed a G-VaR based on sublinear expectation, which is specifically for the condition of variance uncertainty. Besides, we did not find results of calculating the worst-case Value-at-Risk from the empirical data.\par
	In this paper, we proposed a special model uncertainty market model to simply the $\mathcal{P}$ to a set contain finite number of probability distributions. We divide $\mathbb{R}$ into $N$ segments, and in every subset $\mathbb{R}_t$, data is i.i.d.. The model uncertainty market model used the structure of the two-layer mixed distribution model. The first-layer components are Gaussian mixture distributions. Different component distributions correspond to different market factors. The weights of components represent the probability of the occurrence of market factors. We don't care about and unable to model the probability of the occurrence of components. Therefore, when we predict the distribution of future data, we can only know the whole of the component distributions (i.e. market factors) from which the data can be generated, but we cannot accurately know the weights of the components (i.e. the exact probability of any market factor occurring). The second-layer components have no financial meaning and they are just parts of a numerical stimulate method. The remainder of the paper is organized as follows. Section 2 briefly reviews the WVaR(worst-case Value-at-Risk) of return series. In Section 3, we propose a special market model to describe model uncertainty. Section 4 reports the method of returns series segmentation. Section 5 present the method of parameters estimation. In Section 6, we show the empirical results of the WVaR for four financial markets. Finally, Section 7, concludes.\par

\section{WVaR (worst-case Value-at-Risk)}
First we consider the certain condition. Let $X$ be the returns of financial assets, it is a random variable in probability space $(\Omega, \mathcal{F}, P)$. 
\begin{definition}[Value at risk]
Given confident level $\alpha\in [0,1]$, the value at risk $VaR_{\alpha}$ at level $\alpha$ of $X$ with distribution $P$ is
\begin{eqnarray}
VaR_{\alpha}(X) = -inf\{x\in \mathbb{R}:P[X\leqslant x]> \alpha\}
\end{eqnarray}
\end{definition}
In uncertain situation,  we have a set of finitely additive probabilities $\mathcal{P}$ and cannot decide precisely which probability $X$ should obey. Therefore, we have WVaR(worst-case Value-at-Risk) under uncertain conditions.
\begin{definition}(WVaR)
Give a real number  $\alpha \in [0,1]$, the $WVaR_{\alpha}$ at level $\alpha$ of $X$ with a set of finitely additive probabilities 
$\mathcal{P}$ is
\begin{eqnarray}
WVaR_\alpha(X) = -\mathop{\inf}\{x\in\mathbb{R}: \mathop{\max}\limits_{P \in \mathcal{P}}P[X \leqslant x] \geqslant \alpha\}.
\end{eqnarray}
\end{definition}
$WVaR_\alpha(X)$ only care about the “worst” distribution, that is, the distribution of the greatest loss. This “worst case” corresponds to the “worst” scenario for financial assets.\par
	In the definition of $WVaR$, the meaning of $\alpha$ has changed compared with $VaR$. Under certain conditions, $\alpha$ is a confidence level, and $VaR$ is a quantile under $\alpha$. Under uncertain conditions, we cannot decide the probability precisely, therefore, we cannot find the quantile precisely. Under this condition, $\alpha$ becomes a `conditional' confidence level. that is, in the case of a `worst' financial scenario, the probability that the value of asset $X$ is not less than $WVaR_\alpha(X)$ is $\alpha$. This change has led to a change in the criteria for evaluating risk measures. As for $VaR$, we can test the accuracy of method by observing whether the ratio of return to break through $VaR_\alpha(X)$ equals $\alpha$. For $WVaR$, this method is obviously no longer applicable. \par
	Moreover, although many literatures, such as Kass\cite{Kaas1984}, Schepper\cite{Schepper2010}, Popescu\cite{Popescu2005} and He\cite{He2010} studied the properties of $WVaR$, there are rare results about the calculation. One important reason is, in uncertain conditions, it's difficult to estimate the parameters of the distributions in the set of finitely additive probabilities $\mathcal{P}$. In the following paper, we proposed a special market model to simply the $\mathcal{P}$ to a set contain finite number of probability distributions. \par

\section{Model uncertainty market model}	
Let $S_t \in \mathcal{R}$ be the financial asset prices series and $\mathbb{S}=\{S_0,S_1,\cdots,S_{n}\}$. Define returns of the assets as 
\begin{eqnarray}
r_t=log(S_t)-log(S_{t-1}),\ \forall S_t \in \mathbb{S},\ t=1,\cdots,n
\end{eqnarray}
we have financial asset returns series $\mathbb{R}=\{r_1,r_2,\cdots,r_{n}\}$. Suppose that the returns series are independent data and have identical distribution in a short period. We can divide $\mathbb{R}$ into $N$ segments, every segment has $n_t,\ t=1,\cdots,N$ data:
\begin{eqnarray}
&&\mathbb{R}_1=\{r_1,\cdots,r_{n_1}\},\\
&&\cdots\\
&&\mathbb{R}_t=\{r_{n_1+\cdots+n_{t-1}+1}, \ \cdots,r_{n_1+\cdots+n_t}\},\\ 
&&\cdots\\ 
&&\mathbb{R}_N=\{r_{n_1+\cdots+n_{N-1}+1}, \ \cdots,r_{n_1+\cdots+n_N}\},
\end{eqnarray}
where $\sum_{i=1}^{N}n_i=N$. Given a subset $\mathbb{R}_t, \ \forall r_i^{(t)}\in \mathbb{R}_t$ are independent and identically distributed(i.i.d.). For every subset $\mathbb{R}_t,\ t=1,\cdots,N$, we assume that returns are generated from mixture distributions:
\begin{eqnarray}
&&\forall r_i^{(1)} \in \mathbb{R}_1, \ r_i^{(1)} \sim f^{1}(r)=\sum_{j=1}^{K_2}\beta^{1}_{j}p_{j}(r|\theta_j),\\
&&\cdots\\
&&\forall r_i^{(t)} \in \mathbb{R}_t, \ r_i^{(t)} \sim f^{t}(r)=\sum_{j=1}^{K_2}\beta^{t}_{j}p_{j}(r|\theta_j),\\
&&\cdots\\
&&\forall r_i^{(N)} \in \mathbb{R}_N, \ r_i^{(N)} \sim f^{N}(r)=\sum_{j=1}^{K_2}\beta^{N}_{j}p_{j}(r|\theta_j),
\end{eqnarray}
where $p_{j}(x|\theta_j)$ is component distribution, $K_2$ is number of component distributions, $\theta_j$ is the parameters of $j^{th}$ component distribution.\par
	Intuitively, the return $r_i$ can be drawn from one of $K_2$ component distributions $p_{j}(x|\theta_j)$ with the probability $\beta^{t}$. This means that the distribution of return $r_i$ is the result of the comprehensive effect of different market factors, where $K_2$ component distributions $p_{j}(x|\theta_j)$ represents $K_2$ market factors and $\beta^{t}_{j}$ represents the probability of the occurrence of $j^{th}$ market factor. For these financial meanings, our method shares ideas with Zangari\cite{Zangari1996}, Venkataraman\cite{Venkataraman1997}, Hull\cite{Hull1998}, Zhang\cite{Zhang2005}, Billio\cite{Billio2000} and Kawata\cite{Kawata2007}. Within one subset $\mathbb{R}_t$, returns are identically distributed. This setting corresponds to a reasonable assumption that market environment remains static in a short or relative long period. \par
	However, with time goes by, the change of market environment causes the probability of the occurrence of market factors changes. Such kind of change, we assume that, is difficult or even impossible to model because of the complexity of market. On the contrary, the number and type of market factors will not change. That is to say that the change in the distribution of returns is the result of an interaction and trade-off between different market factors, rather than a dramatic change in the number and type of market factors. Therefore, between two different periods $\mathbb{R}_t, \ \mathbb{R}_s$, the distributions of returns $r_i$ changes but the component distributions are unchanged:
\begin{eqnarray}
&&\forall r_i^{(t)} \in \mathbb{R}_t, \ r_i^{(t)} \sim f^{t}(r)=\sum_{j=1}^{K_2}\beta^{t}_{j}p_{j}(r|\theta_j),\\
&&\forall r_i^{(s)} \in \mathbb{R}_s, \ r_i^{(s)} \sim f^{s}(r)=\sum_{j=1}^{K_2}\beta^{s}_{j}p_{j}(r|\theta_j),
\end{eqnarray}	\par	
	Although Zangari\cite{Zangari1996}, Venkataraman\cite{Venkataraman1997}, Hull\cite{Hull1998} and Zhang\cite{Zhang2005} used mixture distribution model, the probability of the occurrence of components is unchanged. Therefore they are certainty probability models. Billio\cite{Billio2000} and Kawata\cite{Kawata2007} modeled return series using HMM(Hidden Markov Model). The HMM model gives the components a certain probability of occurrence by transition probability, although the probability of occurrence is changed at any time. Hence they are also certainty probability models. Different from the above methods, we don't care about and unable to model the probability of the occurrence of components. Therefore, when we predict the distribution of future data, we can only know the whole of the component distributions (i.e. market factors) from which the data can be generated, but we cannot accurately know the weights of the components (i.e. the exact probability of any market factor occurring). That is to say, there is model uncertainty.\par
	 Therefore, the WVaR under uncertain conditions is:
\begin{eqnarray}
WVaR_\alpha(X) = -\mathop{\inf}\{x\in\mathbb{R}: \mathop{\max}P_{p_j}[X \leqslant x] \geqslant \alpha\,\ j=1,2,\cdots,K_2\}.
\end{eqnarray}\par
Different from the GMM(Gaussian mixture model) used by Zangari\cite{Zangari1996}, Venkataraman\cite{Venkataraman1997}, Hull\cite{Hull1998} and Zhang\cite{Zhang2005}, In this paper, the component distributions $p_{j}(r|\theta_j)$ are also mixture distributions:
\begin{eqnarray}
&&p_1(r|\theta_1)=\sum_{i=1}^{K_{1,1}}\alpha_{1,i}N(r|\mu_{1,i},\sigma_{1,i}^2),\\
&&\cdots\\
&&p_j(r|\theta_j)=\sum_{i=1}^{K_{1,j}}\alpha_{j,i}N(r|\mu_{j,i},\sigma_{j,i}^2),\\
&&\cdots\\
&&p_{K_2}(r|\theta_{K_2})=\sum_{i=1}^{K_{1,{K_2}}}\alpha_{{K_2},i}N(r|\mu_{{K_2},i},\sigma_{{K_2},i}^2).
\end{eqnarray}	
Therefore, $r_s$ are generated from a two-layer Gaussian mixture model:
\begin{eqnarray}
&&\forall r_s^{(1)} \in \mathbb{R}_1, \ r_s^{(1)} \sim f^{1}(r)=\sum_{j=1}^{K_2}\beta^{1}_{j}\sum_{i=1}^{K_{1,j}}\alpha_{j,i}N(r|\mu_{j,i},\sigma_{j,i}^2),\\
&&\cdots\\
&&\forall r_s^{(t)} \in \mathbb{R}_t, \ r_s^{(t)} \sim f^{t}(r)=\sum_{j=1}^{K_2}\beta^{t}_{j}\sum_{i=1}^{K_{1,j}}\alpha_{j,i}N(r|\mu_{j,i},\sigma_{j,i}^2),\\
&&\cdots\\
&&\forall r_s^{(N)} \in \mathbb{R}_N, \ r_s^{(N)} \sim f^{N}(r)=\sum_{j=1}^{K_2}\beta^{N}_{j}\sum_{i=1}^{K_{1,j}}\alpha_{j,i}N(r|\mu_{j,i},\sigma_{j,i}^2),
\end{eqnarray}
Admittedly the mixture model with components as Gaussian mixture distributions can be rewritten as a general Gaussian mixture model, for example:
\begin{eqnarray}
\forall r_s^{(t)} \in \mathbb{R}_t, \ r_s^{(t)} \sim f^{t}(r)&=&\sum_{j=1}^{K_2}\beta^{t}_{j}\sum_{i=1}^{K_{1,j}}\alpha_{j,i}N(r|\mu_{j,i},\sigma_{j,i}^2),\\
		&=&\sum_{j=1}^{K_2}\sum_{i=1}^{K_{1,j}}\gamma_{j,i}^{t}N(r|\mu_{j,i},\sigma_{j,i}^2), \ \gamma_{j,i}^{t}=\beta^{t}_{j}\cdot \alpha_{j,i},
\end{eqnarray}
but the first-layer components $p_{j}(r|\theta_j)$ and the second-layer components $N(r|\mu_{j,i},\sigma_{j,i}^2)$ have distinct different meanings: $p_{j}(r|\theta_j)$ represents $j^{th}$ market factor. But $N(r|\mu_{j,i},\sigma_{j,i}^2)$ not have any financial meaning, it is just a part of a numerical stimulate method. This difference is particularly important in calculating UVaR. Reviewing
\begin{eqnarray}
UVaR_\alpha(X) = -\mathop{\inf}\{x\in \mathbb{R}: \mathop{\max}\limits_{P \in \mathcal{P}}P[X \leqslant x] \geqslant \alpha\},
\end{eqnarray}
what we want to do is to find out the worst distribution corresponded to the worst scenario. Using a one-layer Gaussian mixture model with more components rather than a two-layer Gaussian mixture model will make the UVaR overestimated. For example, for a two-layer Gaussian mixture model:
\begin{eqnarray}
\sum_{j=1}^{K_2}\beta^{t}_{j}\sum_{i=1}^{K_{1,j}}\alpha_{j,i}N(r|\mu_{j,i},\sigma_{j,i}^2),
\end{eqnarray}
suppose that the worst distribution is:
\begin{eqnarray}
p_{\hat{j}}(r|\theta_{\hat{j}})=\sum_{i=1}^{K_{1,{\hat{j}}}}\alpha_{ \hat{j} , i }N(r|\mu_{ \hat{j} , i },\sigma_{ \hat{j} , i }^2),
\end{eqnarray}
then, UVaR is:
\begin{eqnarray}
UVaR_{\alpha,p_{\hat{j}}}(X) = -inf\{x\in \mathbb{R}:P_{p_{\hat{j}}}[X\leqslant x]> \alpha\}.
\end{eqnarray}
But for a one-layer Gaussian mixture model with more components
\begin{eqnarray}
\sum_{j=1}^{K_2}\sum_{i=1}^{K_{1,j}}\gamma_{j,i}^{t}N(r|\mu_{j,i},\sigma_{j,i}^2), \ \gamma_{j,i}^{t}=\beta^{t}_{j}\cdot \alpha_{j,i},
\end{eqnarray}
if the ``worst'' distribution:
\begin{eqnarray}
N(r|\bar{\mu},\bar{\sigma}^2) \in \{ N(r|\mu_{ \hat{j} , i },\sigma_{ \hat{j} , i }^2), \ i=1,\cdots,K_{1,{\hat{j}}} \},
\end{eqnarray}
then,
\begin{eqnarray}
UVaR_{\alpha,p_{\hat{j}}}(X)  \leqslant  UVaR_{\alpha, N(r|\bar{\theta})}(X)  = -inf\{x\in \mathbb{R}:P_{N(r|\bar{\theta})}[X\leqslant x]> \alpha\}.
\end{eqnarray}
Moreover, for different returns data, we obviously can't guarantee that:
\begin{eqnarray}
P[N(r|\bar{\mu},\bar{\sigma}^2) \in \{ N(r|\mu_{ \hat{j} , i },\sigma_{ \hat{j} , i }^2), \ i=1,\cdots,K_{1,{\hat{j}}} \}]=0.
\end{eqnarray}
Therefore, using two-layer Gaussian mixture model is reasonable.
\section{Returns series segmentation}
	 As we mentioned in Section 3, we need to divide $\mathbb{R}$ into $N$ segments. Within one subset $\mathbb{R}_t$, $\forall r_i^{(t)}\in \mathbb{R}_t$ are independent and identically distributed(i.i.d.). In this paper, we use Kernel-based detection method to divide $\mathbb{R}$. A kernel-based method has been proposed by Harchaoui\cite{Harchaoui2007} to perform change point detection in a non-parametric setting. Truong\cite{Truong2018} gave a good review about the relative methods. As described by Truong\cite{Truong2018}: to that end, the original series $\mathbb{y}=\{y_1,y_2,\cdots,y_{T}\}$ is mapped onto a reproducing Hilbert space (rkhs) $\mathcal{H}$ associated with a user-defined kernel function $k(\cdot,\cdot):\mathbb{R}^d \times \mathbb{R}^d \to \mathbb{R}$. The mapping function $\phi:\mathbb{R}\to\mathcal{H}$ onto this rkhs is implicitly defined by $\phi(y_t)=k(y_t,\cdot)\in\mathcal{H}$, resulting in the following inner-product and norm:
\begin{eqnarray} 
&&k(y_s,y_t)=\langle \phi(y_s) | \phi(y_t) \rangle_{\mathcal{H}},\\  
&&k(y_t,y_t)=||\phi(y_t)||^2_{\mathcal{H}}
\end{eqnarray}	 
for any samples $y_s,y_t\in \mathbb{R}^d$. The associated cost function, denoted $c_{kernel}$, is defined as follows. 
\begin{eqnarray} 
c_{kernel}(y_{a\cdots b}):=\sum_{t=a+1}^{b}|| \phi(y_t)- \bar{\mu}_{a\cdots b}||^2_{\mathcal{H}}
\end{eqnarray}	
where $y_{a\cdots b}=\{ y_t\}_{t=a+1}^{b}$, $\bar{\mu}_{a\cdots b}\in \mathcal{H}$ is the empirical mean of the series $\{ \phi(y_t)\}_{t=a+1}^{b}$. Indeed, after simple algebraic manipulations, $c_{kernel}(y_{a\cdots b})$ can be rewritten as follows:	
\begin{eqnarray} 
c_{kernel}(y_{a\cdots b}):=\sum_{t=a+1}^{b}k(y_t,y_t)-\frac{1}{b-a}\sum_{t,s=a+1}^{b}k(y_s,y_t).
\end{eqnarray}	
The cost function $c_{kernel}$ can be combined with any kernel to accommodate various types of data. In this paper, we use the Gaussian kernel:
\begin{eqnarray}
k(x,y)=\exp(-\gamma||x-y||^2)
\end{eqnarray}	
with $x,y\in\mathbb{R}^d$ and $\gamma >0$ is the so-called bandwidth parameter. The associated cost function, denoted $c_{rbf}$, is defined as follows:
\begin{eqnarray} 
c_{rbf}(y_{a\cdots b}):=(b-a)-\frac{1}{b-a}\sum_{t,s=a+1}^{b}\exp(-\gamma||x-y||^2)
\end{eqnarray}		
where $\gamma >0$ is the so-called bandwidth parameter.\par
	As described by Truong\cite{Truong2018}: Denote $\mathcal{T}=\{t_1,t_2,\cdots \} \subset \{ 1,\cdots,T\}$	and
\begin{eqnarray} 
V(\mathcal{T}):=\sum_{k=0}^Kc(y_{t_k\cdots t_{k+1}})	
\end{eqnarray}		
where $c(\cdot)$ is a cost function. The change point detection problem with an unknown number of change points consists in solving the following discrete optimization problem	
\begin{eqnarray}
\min_{\mathcal{T}}V(\mathcal{T})+pen(\mathcal{T})
\end{eqnarray}	
where $pen(\mathcal{T})$ is an appropriate measure of the complexity of a segmentation $\mathcal{T}$. Truong\cite{Truong2018} also introduce several penalty functions. In this paper, we use linear penalties which are linear functions of the number of change points, meaning that:
\begin{eqnarray}
pen(\mathcal{T})=\beta|\mathcal{T}|
\end{eqnarray}	
where $\beta>0$ is a smoothing parameter, and $|\mathcal{T}|$ is cardinal of $\mathcal{T}$. In this paper, we used $ruptures$, which is  a Python scientific library provided by Truong\cite{TruongOnline}, to divide $\mathbb{R}$.

\section{Model parameters estimating}
	In this paper, we use EM algorithm to estimate the parameters.  For simplicity, let $K_{1,1}=K_{1,2}=\cdots=K_{1,K_2}=K_1$.  Denote the latent variable
\begin{eqnarray}
\eta_{s,j,i}=\left\{
\begin{aligned}
& 1,\ the\ s^{th}\ observation\ is\ generated\ from\ the\ i^{th}\ component\ of\ the\ j^{th}\ component,  \\
& 0,\ or,
\end{aligned} \right.
\end{eqnarray}	
where $j=1,2,\cdots,K_2$ and $i=1,2,\cdots,K_1$. Denote 
\begin{eqnarray}
&&\hat{n}^{(t)}_{st}=n_1+\cdots+n_{t-1}+1,\\
&&\hat{n}^{(t)}_{en}=n_1+\cdots+n_{t},
\end{eqnarray}	
where $t=1,\cdots,N$. Because the observable variable is $\{r_s\}_{s=1}^{n}$, the complete variable is 
\begin{eqnarray}
(r_s, \eta_{s,1,1},\cdots, \eta_{s,1,K_1}, \eta_{s,2,1},\cdots, \eta_{s,2,K_1},\cdots,\eta_{s,K_2,1},\cdots, \eta_{s,K_2,K_1}),\ s=1,\cdots, n.
\end{eqnarray}
Then, given a subset $\mathbb{R}_t$, the likelihood function of complete variable is
\begin{eqnarray}
P(r,\eta|\theta)&=& \prod_{s=\hat{n}^{(t)}_{st}}^{\hat{n}^{(t)}_{en}}P(r_s, \eta_{s,1,1},\cdots, \eta_{s,1,K_1},\cdots,\eta_{s,K_2,1},\cdots, \eta_{s,K_2,K_1}|\theta)\\
		      &=& \prod_{s=\hat{n}^{(t)}_{st}}^{\hat{n}^{(t)}_{en}} \prod_{j=1}^{K_2} \prod_{i=1}^{K_1} [ \beta_{j}^{(t)} \alpha_{j,i} \phi(r_s | \theta_{j,i}) ]^{{\eta}_{s,j,i}}\\
		      &=& \prod_{j=1}^{K_2} {\beta_{j}^{(t)}}^{ \sum_{i=1}^{K_1} \sum_{ s=\hat{n}^{(t)}_{st} }^{ \hat{n}^{(t)}_{en} } \eta_{s,j,i} } \prod_{i=1}^{K_1} { \alpha_{j,i} }^{ \sum_{ s=\hat{n}^{(t)}_{st} }^{ \hat{n}^{(t)}_{en} } \eta_{s,j,i} } \prod_{s=\hat{n}^{(t)}_{st}}^{\hat{n}^{(t)}_{en}}  [ \phi(r_s | \theta_{j,i}) ]^{{\eta}_{s,j,i}}.
\end{eqnarray}
Denote
\begin{eqnarray}
n_{j,i}^{t} := \sum_{ s=\hat{n}^{(t)}_{st} }^{ \hat{n}^{(t)}_{en}} {\eta}_{s,j,i} \ \ and \ \ n_{j,\cdot}^{t} := \sum_{i=1}^{K_1} \sum_{ s=\hat{n}^{(t)}_{st} }^{ \hat{n}^{(t)}_{en}}{\eta}_{s,j,i}
\end{eqnarray}
then we have 
\begin{eqnarray}
\sum_{i=1}^{K_1}n_{j,i}^{t}=n_{j,\cdot}^{t}\ \ , \ \ \sum_{j=1}^{K_2}n_{j,\cdot}^{t}=n_t\ \ and \ \ \sum_{t=1}^{N}n_t=N.
\end{eqnarray}
Therefore, 
\begin{eqnarray}
P(r,\eta|\theta)= \prod_{j=1}^{K_2} {\beta_{j}^{(t)}}^{ n_{j,\cdot}^{t} } \prod_{i=1}^{K_1} { \alpha_{j,i} }^{ n_{j,i}^{t} } \prod_{s=\hat{n}^{(t)}_{st}}^{\hat{n}^{(t)}_{en}}  [ \phi(r_s | \theta_{j,i}) ]^{{\eta}_{s,j,i}}.
\end{eqnarray}
For the whole series $\mathbb{R}$, the likelihood function of complete variable is
\begin{eqnarray}
P(r,\eta|\theta)= \prod_{t=1}^{N} \prod_{j=1}^{K_2} {\beta_{j}^{(t)}}^{ n_{j,\cdot}^{t} } \prod_{i=1}^{K_1} { \alpha_{j,i} }^{ n_{j,i}^{t} } \prod_{s=\hat{n}^{(t)}_{st}}^{\hat{n}^{(t)}_{en}}  [ \phi(r_s | \theta_{j,i}) ]^{{\eta}_{s,j,i}}.
\end{eqnarray}
The log-likelihood function is
\begin{eqnarray}
\log P(r,\eta|\theta)&=& \sum_{t=1}^{N} \bigg\{ \sum_{j=1}^{K_2} \Big\{ \log( {\beta_{j}^{(t)}} ) n_{j,\cdot}^{t} + \sum_{i=1}^{K_1} \big\{ \log( \alpha_{j,i} ) n_{j,i}^{t} \nonumber \\
		            &+& \sum_{s=\hat{n}^{(t)}_{st}}^{\hat{n}^{(t)}_{en}}  \big[ \log(\frac{1}{\sqrt{2\pi}}) - \log(\sigma_{j,i}) - \frac{(r_s-\mu_{j,i})^2}{2\sigma_{j,i}^{2}} \big] {\eta}_{s,j,i}  \big\} \Big\} \bigg\}.
\end{eqnarray}\par
\textbf{E} \textbf{Step}:\par
	Denote $\theta^{(i)}$ is the parameter obtained in the $i^{th}$ iteration, and
\begin{eqnarray}
Q(\theta, \theta^{(i)}) &=& \mathbb{E}_{\theta^{(i)}}[ \log P(r,\eta|\theta) ]\\
				&=& \sum_{t=1}^{N} \bigg\{ \sum_{j=1}^{K_2} \Big\{ \log( {\beta_{j}^{(t)}} ) \big[ \sum_{i=1}^{K_1} \sum_{ s=\hat{n}^{(t)}_{st} }^{ \hat{n}^{(t)}_{en}} \mathbb{E}_{\theta^{(i)}}[{\eta}_{s,j,i}] \big] \nonumber \\
				&+& \sum_{i=1}^{K_1} \big\{ \log( \alpha_{j,i} ) \big[ \sum_{ s=\hat{n}^{(t)}_{st} }^{ \hat{n}^{(t)}_{en}} \mathbb{E}_{\theta^{(i)}}[{\eta}_{s,j,i}] \big] \nonumber \\
				&+& \sum_{s=\hat{n}^{(t)}_{st}}^{\hat{n}^{(t)}_{en}}  \big[ \log(\frac{1}{\sqrt{2\pi}}) - \log(\sigma_{j,i}) - \frac{(r_s-\mu_{j,i})^2}{2\sigma_{j,i}^{2}} \big] \mathbb{E}_{\theta^{(i)}}[{\eta}_{s,j,i}]  \big\} \Big\} \bigg\}.
\end{eqnarray}	
Then
\begin{eqnarray}
\hat{{\eta}}_{s,j,i}&:=& \mathbb{E}_{\theta^{(i)}}[{\eta}_{s,j,i}] \nonumber \\ 
			&=& P({\eta}_{s,j,i}=1 | r_s, \theta^{(i)}) \nonumber \\
			&=& \frac{ P( r_s | {\eta}_{s,j,i}=1,\theta^{(i)}) P( {\eta}_{s,j,i}=1 | \theta^{(i)}) }{P( r_s | \theta^{(i)})} \nonumber \\
			&=& \frac{ \phi( r_s | \theta_{j,i}^{(i)} ) {\beta_{j}^{(t)}}^{(i)} \alpha_{j,i}^{(i)} }{ \sum_{j=1}^{K_2} \sum_{i=1}^{K_1} \phi( r_s | \theta_{j,i}^{(i)} ) {\beta_{j}^{(t)}}^{(i)} \alpha_{j,i}^{(i)} }.
\end{eqnarray}
We can find that, the parameter $\hat{{\eta}}_{s,j,i}$ should contain $t$. So we have $\hat{{\eta}}_{s,j,i}=\hat{{\eta}}_{s,j,i}^{(t)}$. Denote 
\begin{eqnarray}
\bar{n}_{j,i}^{t} := \sum_{ s=\hat{n}^{(t)}_{st} }^{ \hat{n}^{(t)}_{en}} \hat{{\eta}}_{s,j,i}^{(t)} \ &,& \ \bar{n}_{j,\cdot}^{t} := \sum_{i=1}^{K_1} \sum_{ s=\hat{n}^{(t)}_{st} }^{ \hat{n}^{(t)}_{en}}\hat{{\eta}}_{s,j,i}^{(t)}\ , \\
\bar{n}_t = \sum_{j=1}^{K_2}\bar{n}_{j,\cdot}^{t} \ &,& \ \bar{N}=\sum_{t=1}^{N} \bar{n}_t \ ,
\end{eqnarray}
we have 
\begin{eqnarray}
Q(\theta, \theta^{(i)})&=& \sum_{t=1}^{N} \bigg\{ \sum_{j=1}^{K_2} \Big\{ \log( {\beta_{j}^{(t)}} ) \bar{n}_{j,\cdot}^{t} + \sum_{i=1}^{K_1} \big\{ \log( \alpha_{j,i} ) \bar{n}_{j,i}^{t} \nonumber \\
		            &+& \sum_{s=\hat{n}^{(t)}_{st}}^{\hat{n}^{(t)}_{en}}  \big[ \log(\frac{1}{\sqrt{2\pi}}) - \log(\sigma_{j,i}) - \frac{(r_s-\mu_{j,i})^2}{2\sigma_{j,i}^{2}} \big] \hat{{\eta}}_{s,j,i}^{(t)}  \big\} \Big\} \bigg\}.
\end{eqnarray}\par
\textbf{M} \textbf{Step}:\par
	We need to find $\theta^{(i+1)}$ satisfies
\begin{eqnarray}
\theta^{(i+1)}= \arg\max_{\theta}Q(\theta, \theta^{(i)}).
\end{eqnarray}
We have the iterative formulas
\begin{eqnarray}
\hat{\mu}_{j,i} &=& \frac{ \sum_{t=1}^{N} \sum_{s=\hat{n}^{(t)}_{st}}^{\hat{n}^{(t)}_{en}} \hat{{\eta}}_{s,j,i}^{(t)} r_s }{ \sum_{t=1}^{N} \sum_{s=\hat{n}^{(t)}_{st}}^{\hat{n}^{(t)}_{en}} \hat{{\eta}}_{s,j,i}^{(t)} },\\
\hat{\sigma}_{j,i}^2 &=& \frac{ \sum_{t=1}^{N} \sum_{s=\hat{n}^{(t)}_{st}}^{\hat{n}^{(t)}_{en}} \hat{{\eta}}_{s,j,i}^{(t)} (r_s-\mu_{j,i})^2 }{ \sum_{t=1}^{N} \sum_{s=\hat{n}^{(t)}_{st}}^{\hat{n}^{(t)}_{en}} \hat{{\eta}}_{s,j,i}^{(t)} },\\
\hat{\alpha}_{j,i} &=& \frac{ K_2 }{ \bar{N} } \sum_{t=1}^{N} \bar{n}_{j,i}^{t} \ ,\ \ \hat{\beta}_{j}^{(t)} = \frac{ \bar{n}_{j,\cdot}^{t} }{ \bar{n}_t }.
\end{eqnarray}
\section{Empirical computation}
	We consider two financial markets i.e. Chinese(000001.SH from 1999 to 2018) and American(SPX.GI from 1999 to 2018)  securities market.  First, we use Kernel-based detection method to divide losses series. The smooth parameter $\beta=2.5$. Then we use EM algorithm estimate the parameters. Let $K_2=5, K_1=3$. Finally we calculate WVaR and VaR with tolerance level $\alpha=95\%$. To compare the differences between the two markets, we also calculate the BVaR (best-case Value-at-Risk) :
\begin{definition}(BVaR)
Give a real number  $\alpha \in [0,1]$, the $BVaR_{\alpha}$ at level $\alpha$ of $X$ with a set of finitely additive probabilities 
$\mathcal{P}$ is
\begin{eqnarray}
BVaR_\alpha(X) = -\mathop{\inf}\{x\in\mathbb{R}: \mathop{\min}\limits_{P \in \mathcal{P}}P[X \leqslant x] \geqslant \alpha\}.
\end{eqnarray}
\end{definition}
Obviously, BVaR is Value-at-Risk for the best scenario.\par
\begin{figure}[htbp]
	\centering
	\includegraphics[width=0.9\linewidth]{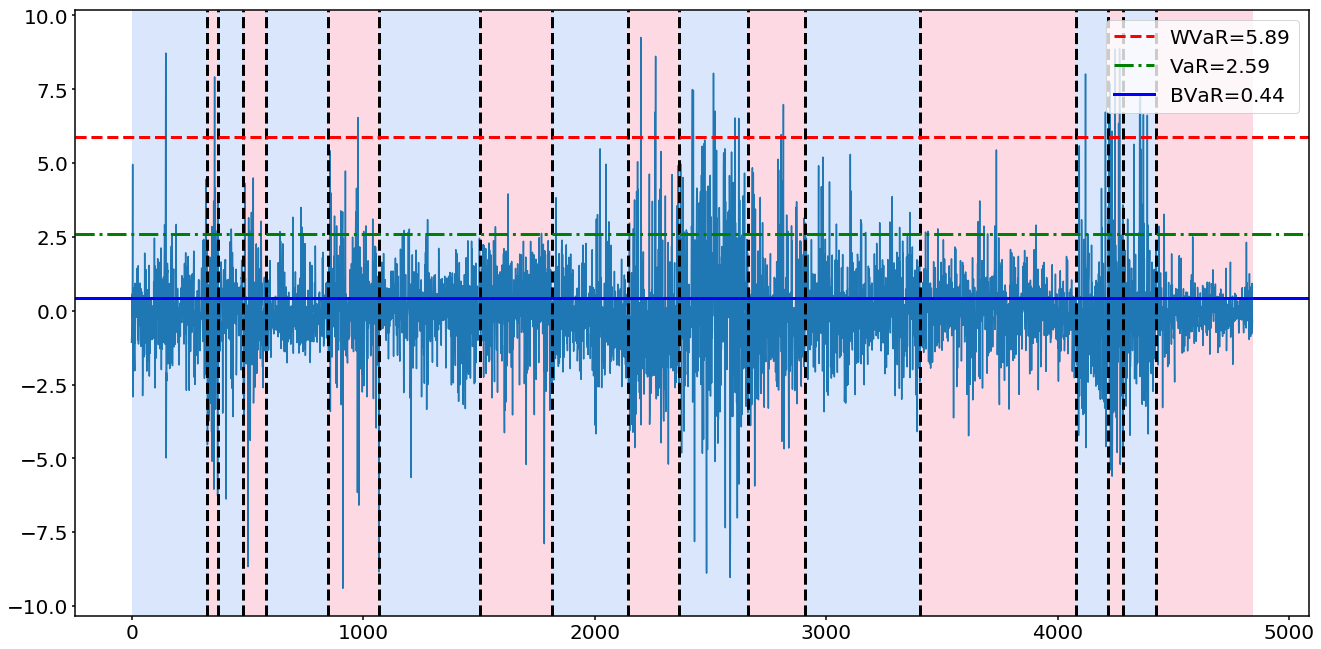} 
	\caption{Chinese(000001.SH) financial market}  
	\label{fig1} 
\end{figure}
\begin{figure}[htbp]
	\centering
	\includegraphics[width=0.9\linewidth]{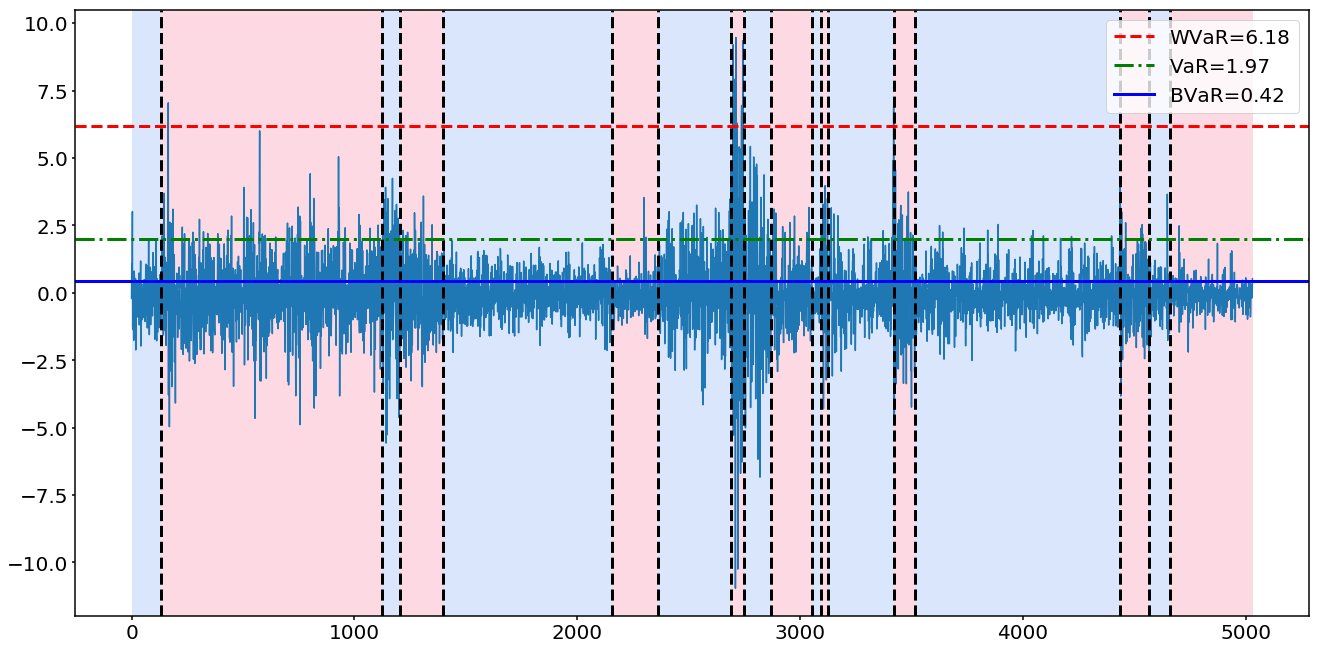} 
	\caption{American(SPX.GI) financial market}  
	\label{fig2} 
\end{figure}

	Figure \ref{fig1} shows that, for Chinese(000001.SH) financial market, we find $17$ change points. $WVaR_{SH}=5.89\%$, $VaR_{SH}=2.59\%$ and $BVaR_{SH}=0.44\%$. Figure \ref{fig2} shows that, for American(SPX.GI) financial market, we find $17$ change points. $WVaR_{SP}=6.18\%$, $VaR_{SP}=1.97\%$ and $BVaR_{SP}=0.42\%$.\par
\par
\par
	Firstly, for the results of losses series segmentation, although two markets both have $17$ change points, the indexes of these points are significant different. For Chinese markets, the occurrence of change points is more uniform. This means that no market condition will last for a long time. But in American markets something is different. The occurrence of change points is more concentrated, and concentrated in the period of high volatility. This means that some ``good'' or ``moderate'' market conditions will last for a relative long time, but in the period of high volatility market condition shifts frequently.\par
	Secondly, from the perspective of risk measures, $WVaR_{SP}>WVaR_{SH}$ which means that American markets has more sever worst-case than Chinese market. American markets and Chinese market have similar best-case because $BVaR_{SH}=0.44\%$ is very near to $BVaR_{SP}=0.42\%$. The interesting thing is, although American markets has more sever worst-case, Chinese market has higher VaR value: $VaR_{SH}=2.59\%>VaR_{SP}=1.97\%$. This fact indicates that, firstly, i.i.d. hypothesis is indeed inappropriate when measuring tail risk. Secondly, for the set of the distributions $\mathcal{P}$ generated by different market factors, the elements of $\mathcal{P}$ of Chinese markets are more concentrated and similar. Yet the elements of $\mathcal{P}$ of American markets tend to perform greater differences.In general, the two markets present different risk characteristics.\par
\begin{figure}[htbp]
	\centering
	\includegraphics[width=0.9\linewidth]{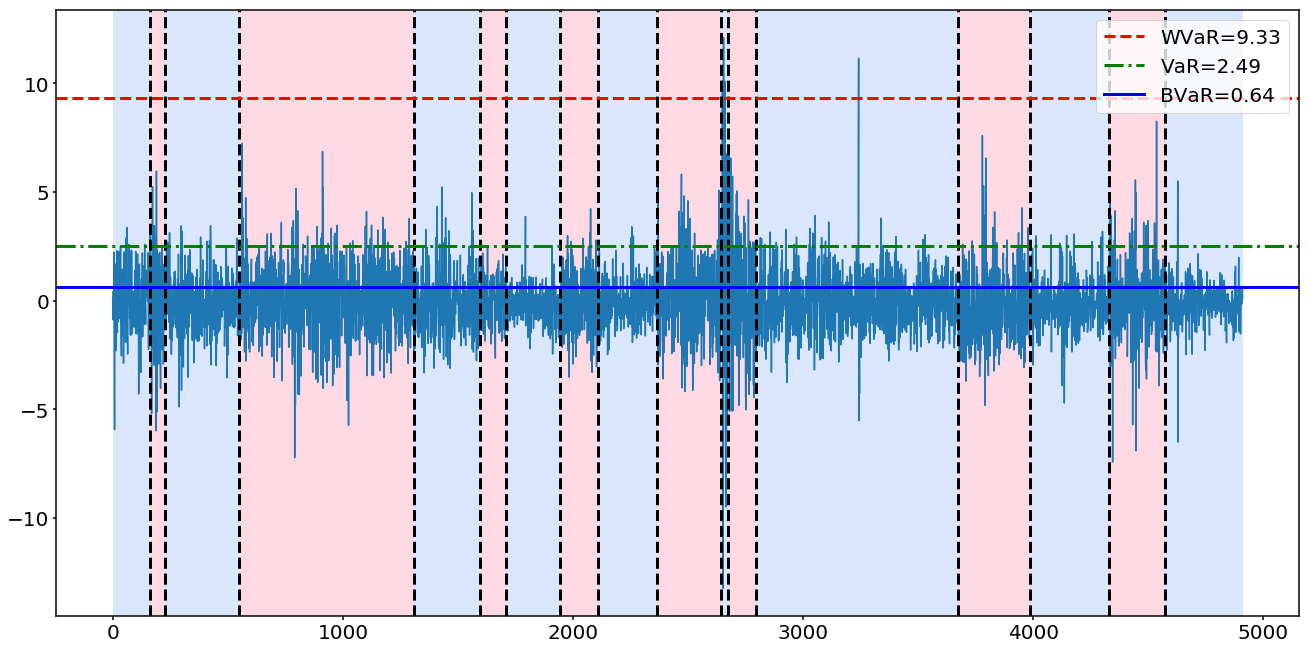} 
	\caption{Japanese(N225.GI) financial market}  
	\label{fig3} 
\end{figure}
\begin{figure}[htbp]
	\centering
	\includegraphics[width=0.9\linewidth]{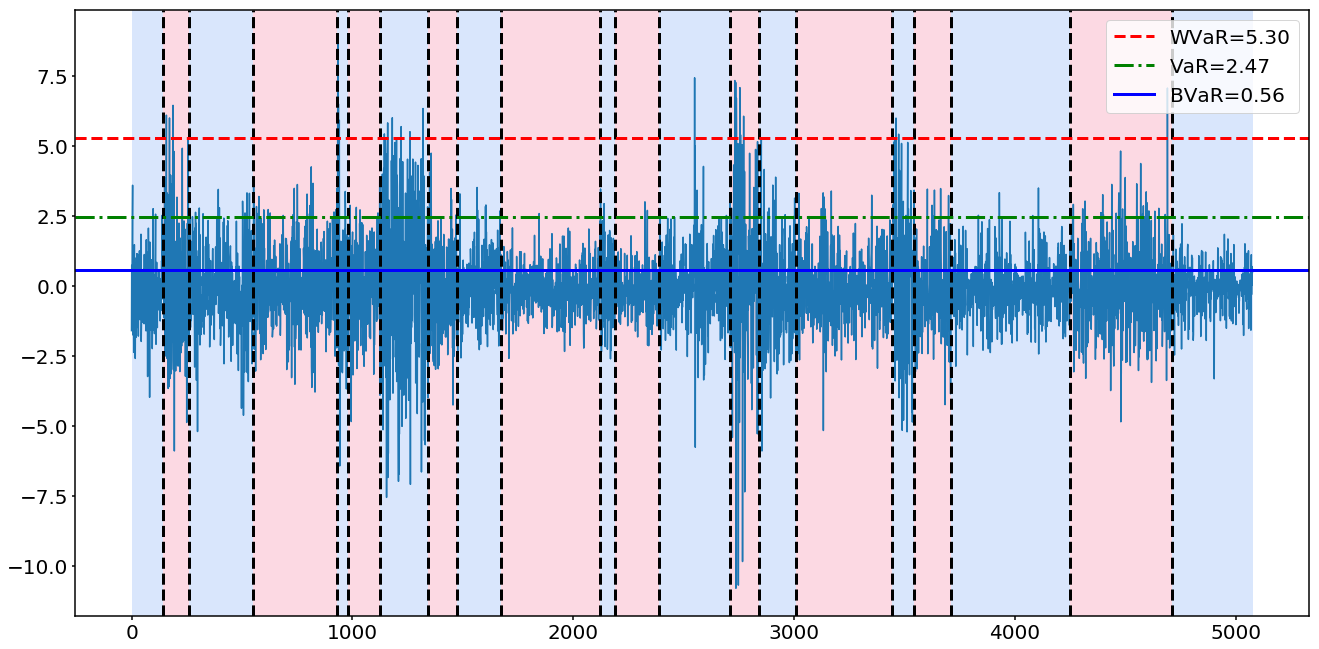} 
	\caption{Germany(GDAXI.GI) financial market}  
	\label{fig4} 
\end{figure}

	In addition, we present the results of Japanese markets (N225.GI from 1999 to 2018) Figure \ref{fig3} and Germany markets (GDAXI.GI from 1999 to 2018) Figure \ref{fig4}. But the analysis of the relevant results is not repeated.\par
\par
\par	 

\section{Conclusion}
	With the rapid development of the financial industry and the frequent emergence of financial crisis, risk management has become an important issue faced by company managers and government regulators. Nowadays, quantification of risky positions held by a financial institution under model uncertainty is of crucial importance from both viewpoints of external regulation and internal management. However, there is no consensus on the sources of model risk. The concept of model uncertainty, sometimes also referred to as model ambiguity. It was argued that data follow not a single distribution but rather a family of distributions. Although we know the the family of models, we cannot precisely decide which one to use. Given the set $\mathcal{P}$, the value of the risk measure $\rho$ varies in a range over the set of all possible models. The largest value in such a range is referred to as a worst-case value, and the corresponding model is called a worst scenario. Although many literatures investigated worst-case Value-at-Risk measures, the implications for empirical data analysis remain rare. \par
	In this paper, we proposed a special model uncertainty market model to simply the $\mathcal{P}$ to a set contain finite number of probability distributions. Suppose that the returns series are independent data and have identical distribution in a short period. We can divide $\mathbb{R}$ into $N$ segments, and in every subset $\mathbb{R}_t$, data is i.i.d.. The model uncertainty market model used the structure of the two-layer mixed distribution model. The first-layer components are Gaussian mixture distributions. Different component distributions correspond to different market factors. The weights of components represent the probability of the occurrence of market factors. We don't care about and unable to model the probability of the occurrence of components. Therefore, when we predict the distribution of future data, we can only know the whole of the component distributions (i.e. market factors) from which the data can be generated, but we cannot accurately know the weights of the components (i.e. the exact probability of any market factor occurring). That is to say, there is model uncertainty. The second-layer components are Gaussian distributions. Actually, the second-layer components have no financial meaning. They are just parts of a numerical stimulate method. For empirical data, firstly we used change point detection method to divide the returns series and then used EM algorithm to estimate the parameters. We calculated VaR, WVaR(worst-case Value-at-Risk) and BVaR(best-case Value-at-Risk) for four financial markets and then analyzed their different performance.

\newpage
\bibliographystyle{unsrt}
\bibliography{ref}

\end{document}